\documentclass[aps, 
               prb, 
               twocolumn, 
               10pt, 
               nofootinbib,
               superscriptaddress,
               amsfonts, 
               amsmath, 
               amssymb,longbibliography]{revtex4-2}

\usepackage[utf8]{inputenc}

\usepackage[T1]{fontenc}
\usepackage{graphicx}
\usepackage{dcolumn}
\usepackage{bm}
\usepackage{xcolor}
\usepackage{lipsum} 
\usepackage{footmisc}
\usepackage{placeins}
\usepackage[colorlinks=true, linkcolor={blue!80!black}, citecolor={blue!80!black}, urlcolor={blue!80!black}]{hyperref}
\usepackage{comment}

\let \revappendix \appendix

\begin{document}

\title{Voltage-tunable Josephson diode at zero applied field}

\author{Sjoerd Telkamp}
\altaffiliation{These authors contributed equally to this work.}
\affiliation{Center for Quantum Devices, Niels Bohr Institute, University of Copenhagen, 2100 Copenhagen, Denmark}
\affiliation{Quantum Center, ETH Z\"urich, 8093 Z\"urich, Switzerland}
\affiliation{Solid State Physics Laboratory, ETH Z\"urich, 8093 Z\"urich, Switzerland}

\author{Junting Zhao}
\altaffiliation{These authors contributed equally to this work.}
\affiliation{Center for Quantum Devices, Niels Bohr Institute, University of Copenhagen, 2100 Copenhagen, Denmark}

\author{Saulius Vaitiekėnas}
\affiliation{Center for Quantum Devices, Niels Bohr Institute, University of Copenhagen, 2100 Copenhagen, Denmark}

\begin{abstract}  
  We report a gate-tunable Josephson diode effect in hybrid nanowire junctions consisting of a spin-orbit-coupled semiconductor core coated with epitaxial ferromagnetic insulator and superconductor shells. The wires display a hysteretic superconducting window as a function of axial magnetic field. In the superconducting regime, the devices exhibit nonreciprocal supercurrent transport, with the diode efficiency showing a strong dependence on back-gate voltage. The effect persists at zero applied magnetic field in a remanent magnetization state following a controlled demagnetization procedure. These findings demonstrate a voltage-controlled Josephson diode in a single junction and suggest a route toward probing intrinsically broken inversion and time-reversal symmetries in hybrid materials.
\end{abstract}
\date{\today}
\maketitle

The Josephson diode effect (JDE)—a nonreciprocal supercurrent in a Josephson junction—has attracted broad interest as a hallmark of broken time-reversal and inversion symmetries in superconducting systems~\cite{Fulton1970, Davydova2022, Zhang2022, Nadeem2023, Nagaosa2023}. 
It has been observed in various platforms, including layered conventional superconductors~\cite{Ando2020}, high-$T_{\rm C}$ superconductors~\cite{Ghosh2024, Volkov2024}, van der Waals heterostructures~\cite{Bauriedl2022}, semiconductor–superconductor hybrids~\cite{Baumgartner2022, Lotfizadeh2024, Reinhardt2024}, supercurrent interferometers~\cite{Souto2022, Matsuo2023, Ciaccia2023, Valentini2024, Coraiola2024,Schrade2024, Banszerus2025}, and thin films with engineered asymmetries~\cite{Hou2023, Castellani2025, Ingla-Aynes2025}. In these systems, symmetry breaking is typically achieved through applied magnetic fields and material composition or device geometry.

Recent efforts have focused on realizing the JDE without external magnetic fields by incorporating magnetic order directly into the materials and device architecture. Field-free JDE has been reported in layered superconductors with magnetic characteristics~\cite{Wu2022, Narita2022, Ma2025,Nagata2025, Li2025}, twisted graphene multilayers~\cite{Scammell2022,Lin2022, Diez-Merida2023, Hu2023}, single magnetic atoms~\cite{Trahms2023}, magnetic domain structures~\cite{Dahir2022, Hess2023, Roig2024,Huang2024}, and ferromagnetic insulators~\cite{Jeon2022}. 
Beyond material-based approaches, nonreciprocal supercurrents have also been engineered by extrinsically breaking time-reversal symmetry using biharmonic microwave drives~\cite{Borgongino2025}, reconfigurable circuit-level asymmetries~\cite{Shi2025}, and multi-terminal systems ~\cite{Chiles2023, Zhang2024, Gupta2023}.
These studies establish a route toward using the JDE to explore intrinsic time-reversal symmetry breaking in superconducting systems.

Semiconductor–superconductor hybrids offer an alternative route to nonreciprocal Josephson transport due to their intrinsic spin-orbit coupling~\cite{Baumgartner2022_2, Legg2022, Turini2022, Costa2023, Costa2023_2, Picoli2023, Fu2024, Meyer2024, Ilic2024, Mori2025, Mazanik2025, Mondal2025}.
In addition, these systems allow independent electrostatic control of carrier density and the proximity effect, enabling direct tuning of junction properties~\cite{Razmadze2023,Maiani2023,Mazur2024, Yan2025}. 
The emergence of spin-triplet correlations in such hybrids may further enhance the superconducting diode effect, particularly in the presence of magnetic inhomogeneity~\cite{Mao2024, Soori2025, Hasan2025}. 
Owing to these properties, the JDE has been proposed and employed as an indirect probe of spin-orbit coupling~\cite{Lotfizadeh2024, Reinhardt2024}.

Here, we report a voltage-tunable Josephson diode effect in the absence of an externally applied magnetic field, realized in a novel triple-hybrid nanowire platform based on semiconducting InAs nanowires proximitized by ferromagnetic insulator (EuS) and superconductor (Al) shell.
We interpret our observations as arising from the interplay between electrostatically controllable spin--orbit coupling, exchange-induced spin splitting, and spatially varying magnetization, thereby establishing a distinct route to the Josephson diode effect.
These ingredients, together with a deterministic demagnetization procedure, enable nonreciprocal supercurrent transport at zero applied field.

\begin{figure}
 
    \centering
    \includegraphics[width = 1\linewidth]{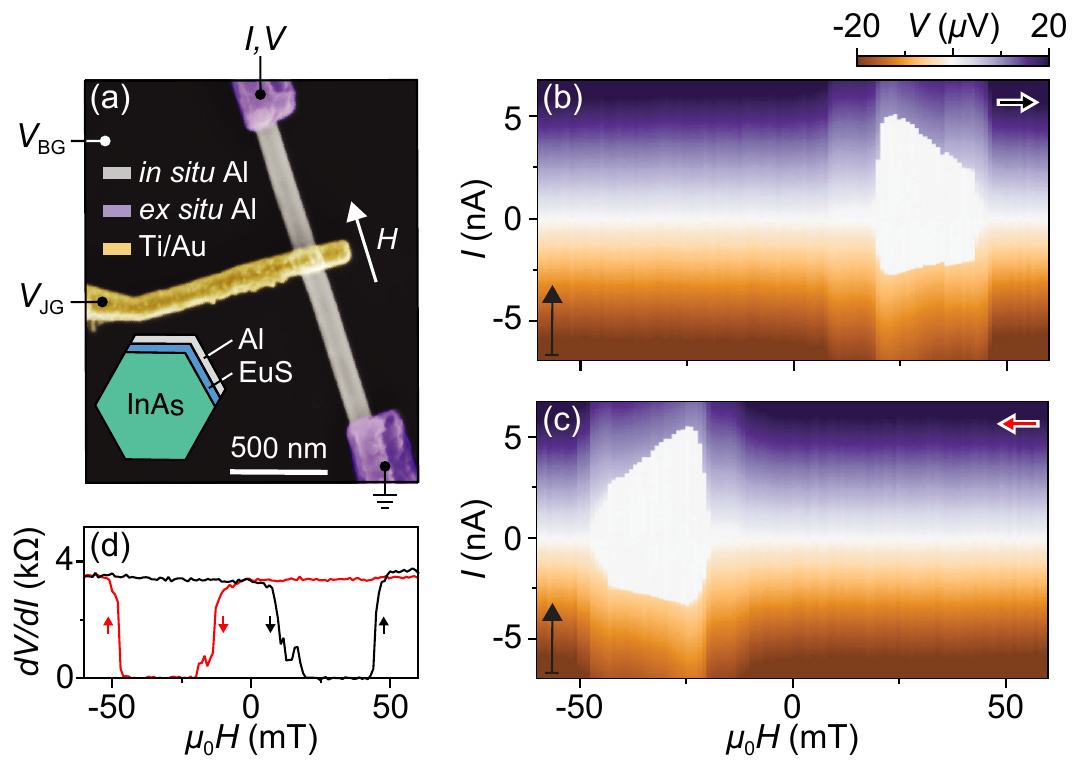}
    
    \caption{ 
    (a) Colorized scanning electron micrograph of a representative device overlaid with the measurement setup. The inset shows schematic wire cross-sections. 
    (b) and (c) Measured dc voltage drop, $V$, across the junction as a function of current bias, $I$, and parallel magnetic field, $H$, swept from (b) negative to positive and (c) positive to negative. Current was ramped from negative to positive in both datasets.
    (d) Differential resistance, $dV/dI$, at zero bias measured as a function of $H$, showing a hysteretic superconducting window.}

    \label{f1}
\end{figure}

The Josephson junctions we investigated are made using hexagonal InAs nanowires grown by molecular beam epitaxy~\cite{Krogstrup2015, Liu2020}, with two of the facets covered \textit{in situ} by fully overlapping epitaxial EuS and Al shells [Fig.~\ref{f1}(a)]---a hybrid platform recently introduced in Ref.~\cite{Zhao2025}. 
Josephson junctions were defined by selectively removing roughly 100~nm of Al, followed by atomic-layer deposition of a thin HfO$_{\rm x}$ dielectric and Ti/Au top-gate metallization for local junction control.
Further fabrication details are provided in Appendix~\ref{app:A}.
To minimize simultaneous tuning of the switching-current amplitude and the higher-harmonic content, the junction-gate voltage was fixed at $V_{\rm JG}=1.9$~V (unless otherwise stated), well away from pinch-off, so that the junction transparency remained approximately constant.
A global back gate was used to tune the charge-carrier density in the nanowire. 
The wires were contacted by \textit{ex-situ} deposited Al leads.
Two nominally identical junctions showing similar behavior have been investigated. 
Measurements were performed using standard low-frequency lock-in techniques in a four-terminal current-bias configuration, in a dilution refrigerator equipped with a vector magnet and a base temperature of 20~mK.
Measurement details are given in Appendix~\ref{app:B}.

The measured nanowires display a hysteretic superconducting window as a function of magnetic field, $H$, applied along the wire axis. 
This is evident in the measured dc voltage drop, $V$, across the junction as a function of current bias, $I$, and field $H$; see Figs.~\ref{f1}(b) and \ref{f1}(c). 
These data were taken after zero-field cooling the sample and subsequently ramping the field to $\mu_0 H = -100$~mT, where the junction exhibits linear $I$--$V$ characteristics, indicating suppressed superconductivity.
Sweeping the field from negative to positive values, the junction remains in the normal state as $H$ crosses zero. 
Nonlinearity emerges in the $I$--$V$ curves around $\mu_0 H = 20$~mT, with a zero-voltage branch centered near $I = 0$, indicating the onset of superconductivity. 
The branch expands rapidly with increasing $H$, reaching approximately $I = 5$~nA at $\mu_0 H = 25$~mT, then gradually shrinks and disappears around $\mu_0 H = 45$~mT, beyond which the junction remains normal. 
The behavior is reversed when sweeping the field from positive to negative, with the superconducting window appearing between $\mu_0 H = -20$ and $-45$~mT.
The superconducting window is further reflected by the evolution of the zero-current differential resistance, $dV/dI$, as a function of $H$, showing a sweep-direction-dependent zero-resistance state; see Fig.~\ref{f1}(d).

This hysteretic behavior can be understood in terms of magnetic domain dynamics in the EuS shell~\cite{Strambini2017, Maiani2023, Zhao2025}. 
At large $H$, EuS is magnetized and the domain size, $d$, exceeds the superconducting coherence length, $\xi$, resulting in a uniform exchange-field-induced spin splitting that suppresses superconductivity. 
As the field approaches the coercive field, $H_{\rm C}$, the shell breaks into a multidomain configuration with $d \lesssim \xi$, yielding a reduced effective spin splitting--averaged over the scale of a Cooper pair--that allows superconductivity to reemerge.

\begin{figure}
 
    \centering
    \includegraphics[width = 1\linewidth]{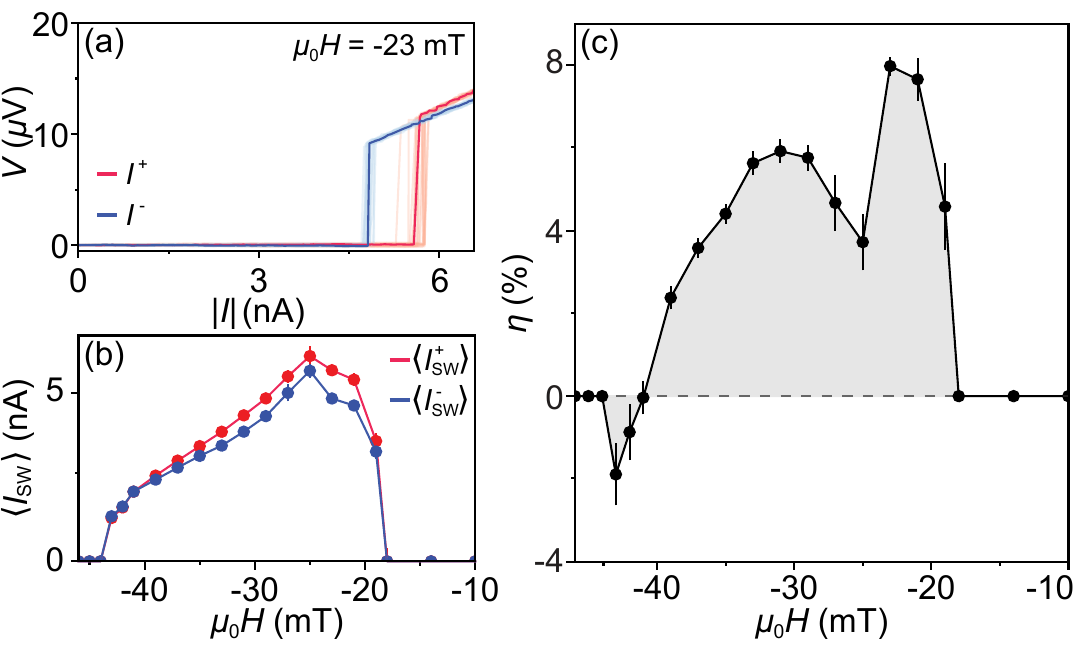}
    
    \caption{
    (a) Measured dc voltage, $V$, as a function of current bias, $I$, taken at fixed magnetic field $\mu_0 H = -23$~mT for 20 repetitions in both sweep directions. The traces closest to the average switching currents $I_{\mathrm{sw}}^{+}$ and $I_{\mathrm{sw}}^{-}$ are highlighted in bold.
    (b) Average switching currents, $\langle I_{\mathrm{sw}}^{\pm} \rangle$, extracted from repeated measurements as in (a) at different magnetic fields, $H$, showing that $\langle I_{\mathrm{sw}}^{-} \rangle$ is systematically smaller than $\langle I_{\mathrm{sw}}^{+} \rangle$ over most of the superconducting window.
    (c) Diode efficiency, $\eta$, as a function of $H$, revealing a finite nonreciprocal response across the superconducting window, with a local minimum around $\mu_0 H = -25$~mT, where $I_{\rm SW}$ is maximal. 
    Error bars indicate uncertainty from the standard error of repeated measurements at each $\mu_0 H$ value.}
    \label{f2}
\end{figure}

We investigate the switching current characteristics of the junction by fixing the magnetic field at $\mu_0 H = -23$~mT, near the maximum of the superconducting window. 
We repeatedly sweep the current bias up and down for 20 cycles, as commonly done to characterize the stochastic switching-current distributions in Josephson-diode measurements~\cite{Ghosh2024, Trahms2023}.
The resulting $I$--$V$ traces reveal a significant and reproducible difference between the positive and negative switching currents, $I_{\mathrm{SW}}^+$ and $I_{\mathrm{SW}}^-$; see Fig.~\ref{f2}(a).
In these measurements, we account for offsets coming from the current source and preamplifier; see Appendix~\ref{app:C}. 
The directional dependence of $I_{\rm SW}$ persists across the entire superconducting window; see Fig.~\ref{f2}(b).
We attribute this asymmetry to the JDE.
The corresponding diode efficiency,
$\eta=\frac{I_{\mathrm{SW}}^+-\left|I_{\mathrm{SW}}^-\right|}{I_{\mathrm{SW}}^++\left|I_{\mathrm{SW}}^-\right|} \times 100 \%$,
peaks at roughly 8\% near the beginning of the superconducting window, exhibits a local minimum at $\mu_0 H = -25$~mT where $I_{\mathrm{SW}}$ is maximal, and changes sign at the high-field edge of the window; see Fig.~\ref{f2}(c). 
The sign reversal is reproducible across multiple field sweeps and is qualitatively consistent with the phenomenological modeling presented in Appendix~\ref{app:I}, where it occurs in the parameter regime in which the harmonic content of the current--phase relation changes and higher harmonics become more pronounced.

Typically, the JDE is realized by breaking time-reversal symmetry with a transverse magnetic field~\cite{Baumgartner2022, Davydova2022}. In our case, the diode effect emerges even though the induced magnetization is expected to be predominantly aligned with the wire axis due to shape anisotropy and the parallel magnetic field used to magnetize the sample~\cite{Vaitiekenas2020, Razmadze2023}. 
At the same time, superconductivity re-emerges only in a narrow field window near the coercive field, pointing to a multidomain regime in which magnetic texture becomes important~\cite{Nandi2026}. 
This suggests that the observed Josephson diode response arises from the interplay of spin--orbit coupling, exchange-induced spin splitting, and a noncollinear magnetic texture near the junction. 
In this picture, structural inversion asymmetry provides spin--orbit coupling, the magnetization from the EuS shell breaks time-reversal symmetry, and the required transverse component arises from the multidomain state near coercivity.
Control measurements on an otherwise similar nonmagnetic Al/InAs junction shown a suppressed diode response under comparable gate and parallel-field conditions; see Appendix~\ref{app:H}.
The phenomenological modeling presented in Appendix~\ref{app:I} reproduces the main qualitative trends of the experiment and supports this interpretation.

We note that reversing the global magnetization direction does not invert the diode polarity. By contrast, magnetizing the wire with a finite off-axis field component shows that the intrinsic diode efficiency can be either enhanced or suppressed depending on the sign of the transverse field component; see Appendix~\ref{app:D}. 
This behavior suggests that the diode effect requires a transverse magnetic component, but that this component is not determined solely by the sign of the global magnetization.
A natural interpretation is that near coercivity the EuS shell forms a multidomain texture~\cite{Nandi2026}, in which local asymmetries at the junction---such as interface pinning, electrostatic inhomogeneity, or geometric asymmetry---together with chirality-selecting interfacial interactions can favor one local transverse orientation~\cite{Kuepferling2023}. 
In this picture, the observed JDE remains tied to the EuS magnetization, while its polarity is controlled by the local magnetic texture together with the structurally and electrostatically defined inversion asymmetry of the device, rather than by the sign of the global magnetization alone.

\begin{figure}
 
    \centering
    \includegraphics[width = 1\linewidth]{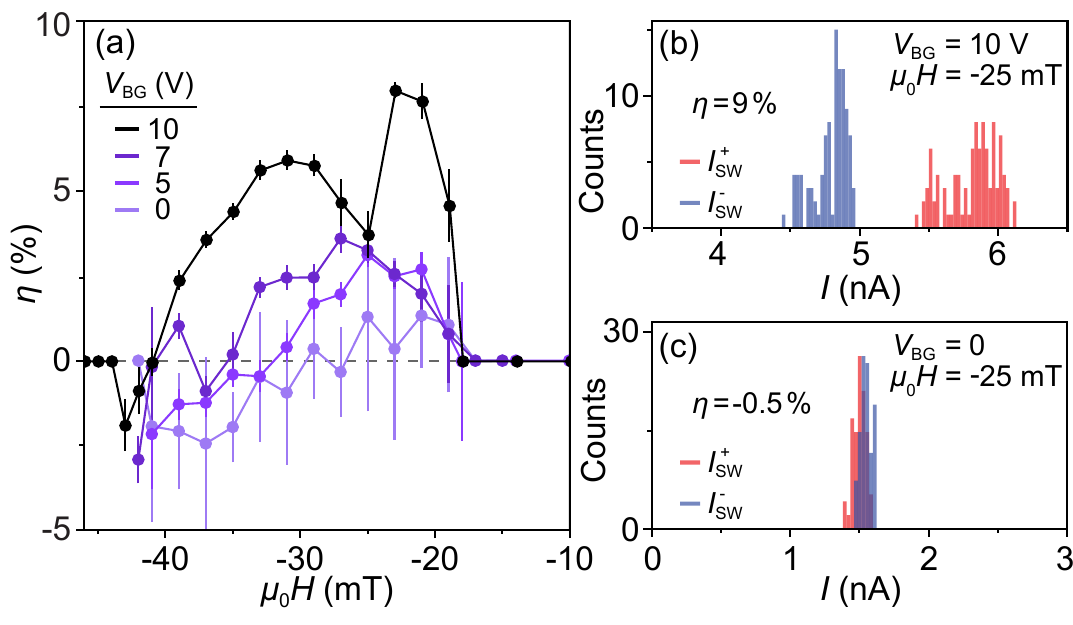}
    
    \caption{
    (a) Diode efficiency, $\eta$, as a function of magnetic field, H, measured for different back-gate voltages, $V_{\rm BG}$, showing a decreasing $\eta$ as $V_{\rm BG}$ is lowered across the superconducting window.  Each point is calculated from 20 current-sweep cycles.
    (b) and (c) Histograms of switching currents $I_{\mathrm{sw}}^{+}$ and $I_{\mathrm{sw}}^{-}$, extracted from separate 100 repetitions at $\mu_0 H = -25$~mT, for (b) $V_{\mathrm{BG}} = 10$~V and (c) $V_{\mathrm{BG}} = 0$. A clear separation between $I_{\mathrm{sw}}^{+}$ and $I_{\mathrm{sw}}^{-}$ can be seen at 10~V, yielding a finite diode efficiency, while the distributions largely overlap at 0~V, resulting in suppressed $\eta$.}
    \label{f3}
\end{figure}

To further assess the role of spin–orbit coupling, we investigate the back-gate voltage dependence of JDE. All measurements discussed so far were performed at $V_{\rm BG} = 10$~V.
We find that the overall diode efficiency across the superconducting window decreases as $V_{\rm BG}$ is lowered; see Fig.~\ref{f3}(a).
At $V_{\rm BG} = 0$~V, the efficiency is effectively suppressed, with most values consistent with $\eta=0$.
This trend is further highlighted by the histograms of $I_{\mathrm{SW}}^{+}$ and $I_{\mathrm{SW}}^{-}$ extracted at $\mu_0 H = -25$~mT; see Figs.~\ref{f3}(b) and \ref{f3}(c).
At $V_{\rm BG} = 10$~V, the two distributions are clearly separated, yielding a diode efficiency of $(9 \pm 0.2)\%$, while at $V_{\rm BG} = 0$~V, they nearly fully overlap, resulting in an efficiency of $(-0.5 \pm 0.7)\%$, consistent with zero. 
We note that the uncertainty at zero gate is dominated by the small $I_{\mathrm{SW}}$ difference entering the error propagation.
Additional measurements at negative back-gate voltage, together with gate-leakage controls, are presented in Appendix~\ref{app:F}. The same qualitative field and gate dependence is reproduced in a second nominally identical junction; see Appendix~\ref{app:G}.

The observed voltage tunability of JDE provides additional support for the interpretation that intrinsic spin–orbit coupling plays a central role in enabling nonreciprocal supercurrent transport.
At a more positive back-gate voltage, the electric field pulls the charge carriers further into the nanowire core away from the shells, enhancing the semiconducting character of the hybrid system~\cite{Vaitiekenas2018, Antipov2018, Mikkelsen2018, Reeg2018, Moor2018, Escribano2021, Escribano2022}.
As a result, the relative contribution of spin–orbit coupling increases compared to the proximity-induced superconductivity~\cite{Zhao2025}.
Conversely, at less positive gate voltages, including $V_{\rm BG}=0$,  a larger fraction of the hybrid wavefunction resides in the superconducting shell, which reduces the effective spin--orbit coupling.
Within this picture, tuning $V_{\rm BG}$ modifies the structural inversion asymmetry experienced by the charge carriers, thereby controlling the strength of the diode effect.
The observed gate dependence is therefore consistent with a role of spin–orbit coupling in the diode response. We note, however, that gate tuning may additionally modify the Josephson energy and thereby the relative importance of thermal phase fluctuations and higher harmonics, potentially contributing to the observed gate dependence.

\begin{figure}
    \centering
    \includegraphics[width = 1\linewidth]{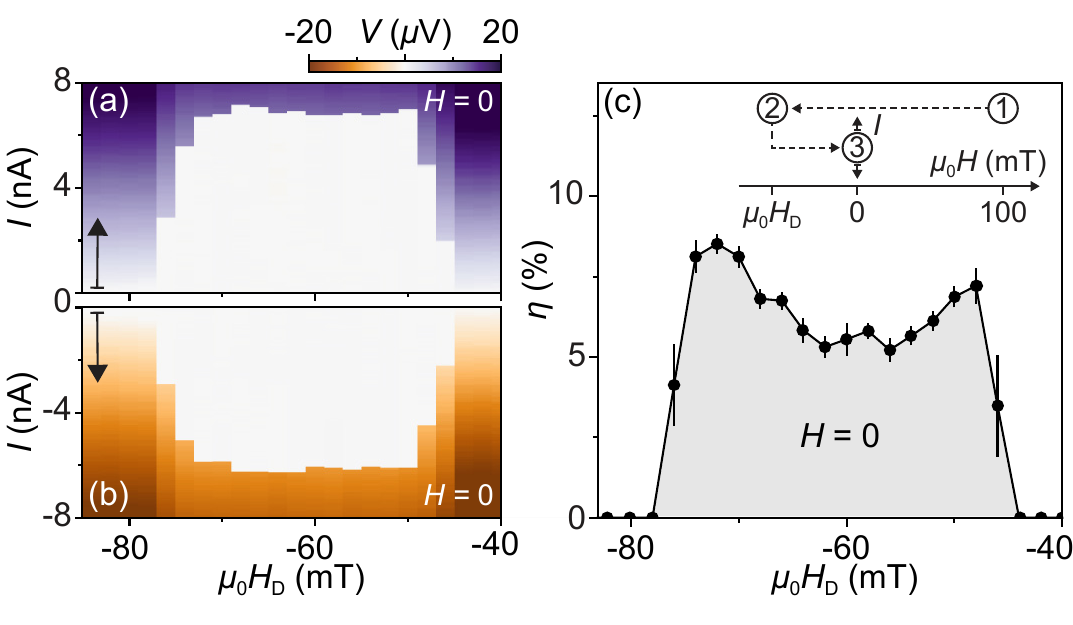}
    \caption{
    (a) and (b) Measured dc voltage, $V$, as a function of current bias, $I$, and demagnetization field, $H_{\rm D}$, for (a) positive and (b) negative current sweeps, starting from $I=0$. Before each measurement, the EuS shell is first magnetized at $\mu_0 H = 100$~mT, followed by sweeping the magnetic field to a given demagnetization value $H_{\rm D}$, and then turning it off before taking $I$--$V$ traces at $H = 0$.
    (c) Diode efficiency, $\eta$, as a function of $H_{\rm D}$, showing a nonmonotonic dependence on demagnetization history, with enhanced efficiency at the boundaries of the superconducting range. The inset shows a schematic illustration of the demagnetization procedure.}
    \label{f4}
\end{figure}

Finally, we demonstrate that JDE in hybrid junctions can be realized at zero applied magnetic field after a controlled demagnetization of the EuS shell (similar to that used in Ref.~\cite{Razmadze2023}), leaving the system in a remanent magnetic state.
The shell is first fully magnetized by applying a positive $\mu_0 H = 100$~mT.
The field is then swept to a negative demagnetizing value $H_{\rm D}$, and subsequently turned off before performing the measurement at $H = 0$. This procedure is carried out at $I = 0$ and repeated for each value of $H_{\rm D}$, followed by current sweeps in the positive or negative direction; see Figs.~\ref{f4}(a) and \ref{f4}(b).

Superconductivity at $\mu_0 H = 0$ is observed for $H_{\rm D}$ values ranging from roughly $-80$ to $-45$~mT.
This can be understood as a partially demagnetized state in which the field reversal does not fully reorient the EuS domains, allowing a remanent magnetization to persist. 
Within this superconducting range, the $I$--$V$ characteristics exhibit nonreciprocal transport.
The corresponding $\eta$ shows a nonmonotonic dependence on the demagnetization history, with maxima near the boundaries of the superconducting range; see Fig.~\ref{f4}(c).
As in the finite-field case shown in Fig.~\ref{f3}, the diode effect in the demagnetized state remains tunable via $V_{\rm BG}$; see Appendix~\ref{app:F}.

Our experiments demonstrate voltage-tunable nonreciprocal supercurrent transport in semiconductor–ferromagnetic insulator–superconductor nanowires. 
The JDE emerges near the coercive field of the magnetic shell—where superconductivity is restored—and persists in the absence of an external magnetic field following a controlled demagnetization procedure. 
The gate-voltage dependence suggests that intrinsic spin–orbit coupling is the dominant inversion symmetry-breaking mechanism, while the time-reversal symmetry is broken by the magnetic shell. 
These findings establish a new scalable materials platform for exploring intrinsic symmetry-breaking mechanisms in hybrid superconducting systems.

\section*{ACKNOWLEDGMENTS}
We thank F.~S.~Bergeret, L.~Casparis, F.~Krizek, A.~Maiani, A.~Manesco, R.~S.~Souto, and W.~Wegscheider for valuable discussions, P.~Krogstrup, Y.~Liu, and C.~S\o rensen for contributions to materials growth, and D.~Razmadze and S.~Upadhyay for nanofabrication.
Research was supported by the Danish National Research Foundation, European Innovation Council [Grant Agreement No. 101115548 (FERROMON)], research grants (Projects No. 43951 and 53097) from VILLUM FONDEN, the Swiss National Science Foundation (SNSF), and the Swiss National Center of Competence in Research Quantum Science and Technology, QSIT.

\section*{DATA AVAILABILITY}
The data used to generate the figures in this work are available in Ref.~\cite{Telkamp2026Zenodo}.

\revappendix

\section{SAMPLE PREPARATION}\label{app:A}

InAs nanowires with hexagonal cross sections were grown using molecular beam epitaxy (MBE) to a length of roughly $10~\mu$m and a diameter of $120$~nm. 
Two of the six facets were coated \textit{in situ} with fully overlapping epitaxial EuS (1~nm) and Al (6 nm) shells.
After growth, individual wires were transferred using a micromanipulator onto a Si chip with 200~nm SiO$_{\rm x}$ capping.
The doped Si substrate also served as a global back gate.
Josephson junctions were formed by selectively etching the \textit{in-situ} Al using a 13\% CSAR 62 resist and a 0.17~N IPA:TMAH developer (AR 300-475, 16:1) for 4~min at room temperature. 
Following native oxide removal by Ar-ion plasma cleaning (25~W, 18~mTorr, 9~min), \textit{ex-situ} Al (190~nm) was deposited to form contacts.
Junction gates were patterned from Ti/Au (5/195~nm) after atomic-layer deposition of HfO$_{\rm x}$ (6~nm) dielectric.

\section{MEASUREMENTS}\label{app:B}

Transport measurements were performed using standard dc and low-frequency ac lock-in techniques in a cryofree dilution refrigerator with a three-axis (1,\,1,\,6)~T vector magnet and a base temperature of 20~mK. 
The dc bias to the device was applied via a load resistor $R_{\rm dc} = 1$~M$\Omega$, whereas the ac bias was applied via a combination of a home-built voltage divider (1:1000) and a load resistor $R_{\rm ac} = 10$~M$\Omega$.
Measurement lines were filtered inside the cryostat using a combination of rf and RC filters, adding a line resistance $R_{\rm Line} = 5.8$~k$\Omega$. 
Gate lines were additionally filtered at room temperature using custom-built 16~Hz low-pass filters.  
Currents were transamplified using an
$I$--$V$ converter with a gain of 10$^6$.
Voltages were preamplified by low-noise preamplifiers with a gain of 10$^3$. 
Four-probe differential resistance measurements were performed using 50~pA  excitation at 18.99~Hz.
For each measurement condition, \(\eta\) is calculated from the mean switching currents, \(\langle I_{\mathrm{SW}}^{+}\rangle\) and \(\langle I_{\mathrm{SW}}^{-}\rangle\). The quoted uncertainties in \(\eta\) are propagated from the statistical uncertainties of these mean values and do not directly correspond to the widths of the underlying switching-current distributions.

\section{CURRENT OFFSET CORRECTION}\label{app:C}

\begin{figure}[t!]
 
    \centering
    \includegraphics[width = \linewidth]{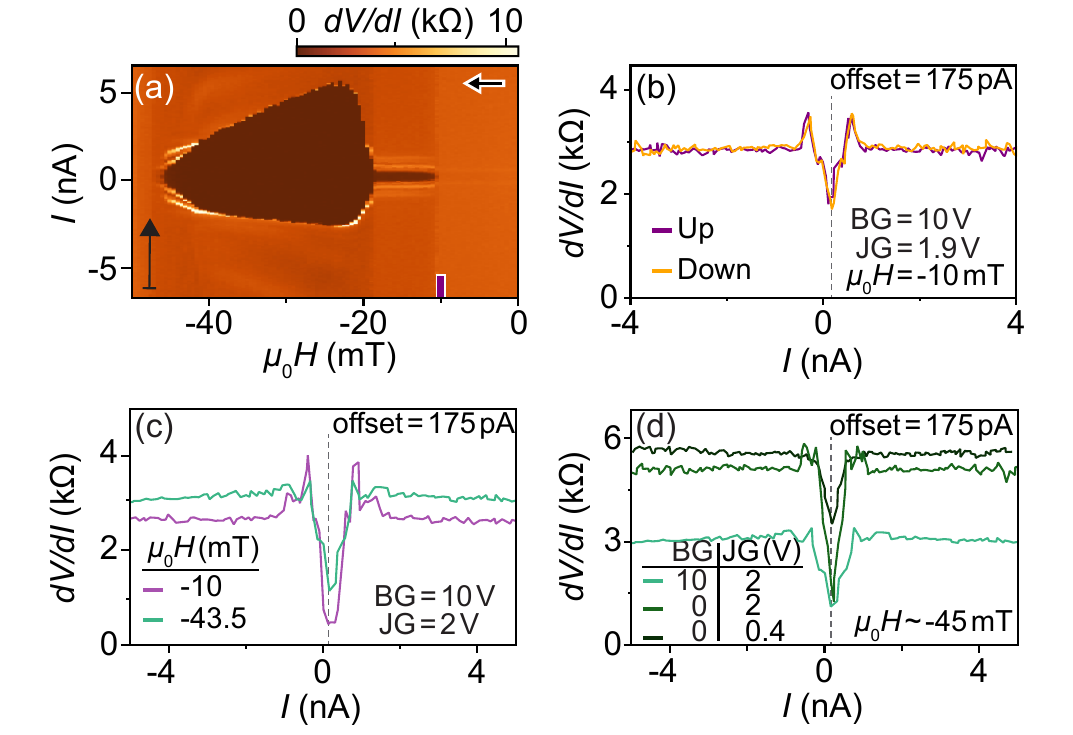}
    
    \caption{(a)~Differential resistance, $dV/dI$, as a function of measured dc current, $I$, and parallel magnetic field, $H$. The arrow indicates the current-sweep direction, capturing the retrapping and switching currents.
    (b) Line traces of $dV/dI$ at $\mu_0H=-10$~mT for upward and downward current sweeps, showing a common resistance minimum at $I=175$~pA, identified as the systematic current offset.
    (c) Traces of $dV/dI$ at different magnetic fields showing that the offset position is field independent.
    (d) Traces of $dV/dI$ for different back- and junction-gate voltages showing that the offset is likewise independent of gate voltage.
    The settings of the junction and backgate voltages, as well as the magnetic field for each trace, are indicated in the corresponding panels. 
    The extracted $175$~pA offset is subtracted from all dc current data.
    }
    \label{f5}
\end{figure}

All dc current and voltage measurements were numerically corrected for systematic offsets introduced by the current source, $I$--$V$ transamplifier, and voltage preamplifier.
The current offset was determined using differential resistance, $dV/dI$, maps measured as a function of measured dc current, $I$, and applied magnetic field, $H$; see Fig.~\ref{f5}(a).
A line cut taken before the fully-developed superconducting regime, at $\mu_0 H = -10$~mT, shows a sharp resistance minimum at $I = 175$~pA in both current sweep directions; see Fig.~\ref{f5}(b).
This value is identified as the current offset and subtracted from all dc current data.
Importantly, the position of this feature remains unchanged upon varying the magnetic field [Fig.~\ref{f5}(c)] and the gate voltages [Fig.~\ref{f5}(d)], demonstrating that the offset is independent of the experimental parameters relevant to the diode measurements.
The dc voltage offset is corrected by subtracting the measured voltage around $I = 0$ for each trace.

\section{THEORETICAL MODELING}\label{app:I}

A finite JDE requires a spin-splitting field with a component transverse to the current flow~\cite{Yokoyama2014, Davydova2022, Meyer2024}. In our devices, shape anisotropy is expected to align the magnetization predominantly along the nanowire axis~\cite{Vaitiekenas2020, Liu2020}. However, in the multidomain regime---which is also the regime where superconductivity re-emerges~\cite{Razmadze2023, Maiani2025}---the magnetization can rotate away from the wire axis in the region separating neighboring domains~\cite{Nandi2026}. The average local magnetic texture near the junction may then be influenced by interface pinning~\cite{Diesch2018}, or more generally by the interplay of a nonuniform electrostatic environment, spin--orbit coupling, and broken interfacial inversion symmetry~\cite{Kuepferling2023}. 
It is also useful to note that in one-dimensional systems, spatially varying spin--orbit coupling in a homogeneous field is gauge equivalent to homogeneous spin--orbit coupling in a spin-splitting field with a spatial magnetic texture~\cite{Kjaergaard2012, Bergeret2014, Maiani2023}.

\begin{figure}[t!]
 
    \centering
    \includegraphics[width = \linewidth]{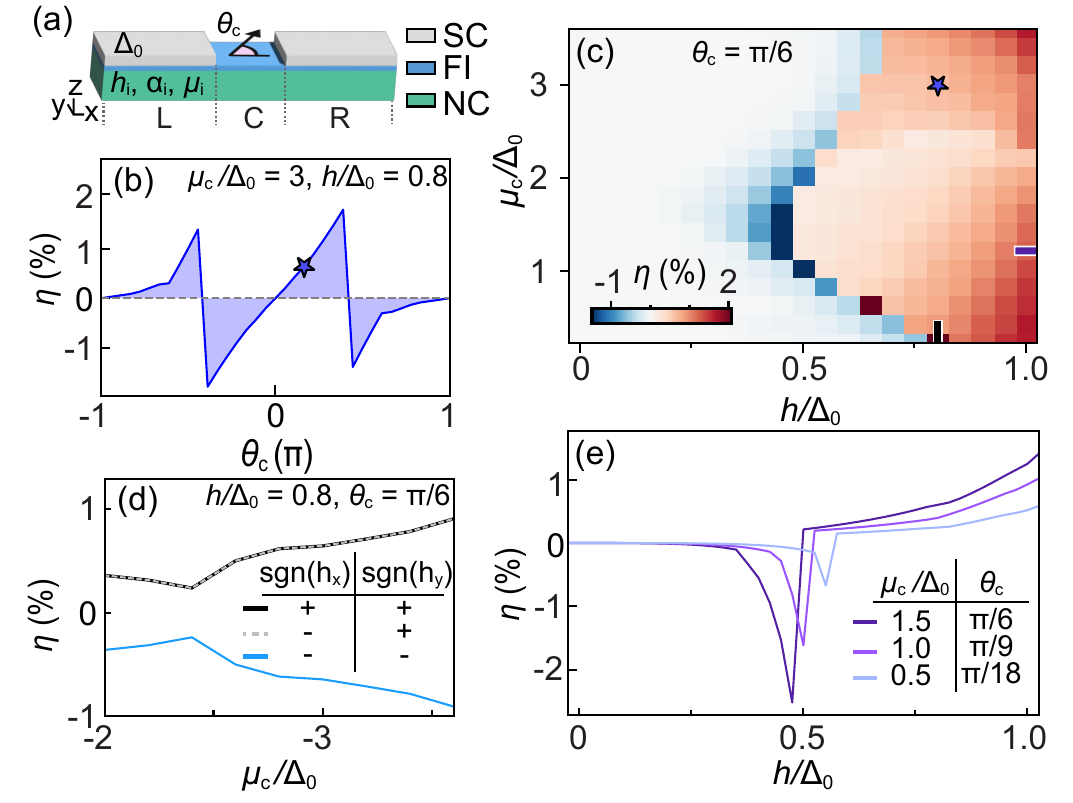}
    
    \caption{
        (a)~Schematic illustration of one-dimensional Josephson junction model adapted from Ref.~\cite{Maiani2023} with left (L), central (C), and right (R) regions, each characterized by the chemical potential $\mu_i$, spin-orbit coupling $\alpha_i$, and spin-splitting field $\mathbf{h}_i$. The spin-splitting field in the leads is taken along the wire, while $\mathbf{h}_{\rm C}$ can be rotated off the wire axis by $\theta_{\rm C}$.
        (b)~Calculated Josephson diode efficiency, $\eta$, as a function of $\theta_{\rm C}$ at fixed $|h|/\Delta_0=0.8$, showing that a finite off-axis component in the central region is required to obtain a finite diode effect.
        (c)~Phase diagram of $\eta(\mu_{\rm C},h)$ as a function of $\mu_{\rm C}$ and the amplitude of $\textbf{h}$ in all three regions with $\theta_{\rm C}=\pi/6$. 
        (d)~$\eta(\mu_{\rm C})$ at $|h|/\Delta_0=0.8$ and $\theta_{\rm C}=\pi/6$ for different sign choices of the axial and transverse field components, illustrating that reversing the axial component (in all three regions) alone does not invert the diode polarity, whereas reversing the transverse component (in the central region) does.
        (e) $\eta(h)$ for selected $(\mu_{\rm C},\theta_{\rm C})$ pairs mimicking back-gate voltage tuning.
        Here $h$ denotes the effective spin-splitting field (exchange) in the model, which is not expected to be proportional to the applied field due to magnetic hysteresis.
        }
        
    \label{f6}
\end{figure}

To get a better qualitative understanding of the observed nonreciprocal response, we use the one-dimensional continuum model introduced in Ref.~\cite{Maiani2023}, with the code provided in Ref.~\cite{MaianiZenodo}.
The model describes a semiconductor--ferromagnetic insulator--superconductor hybrid Josephson junction consisting of left (L) and right (R) superconducting leads and a central (C) junction region; see Fig.~\ref{f6}(a). 
The Hamiltonian used to describe the system contains the usual kinetic-energy term with region-dependent chemical potential, superconducting pairing in the leads, Rashba spin--orbit coupling, and an effective spin-splitting field.
It allows for independent control of the chemical potential, $\mu_i$, spin--orbit coupling, $\alpha_i$, and spin-splitting field, $\mathbf{h}_i$, with $i\in\{L,C,R\}$. 
We note that $\mathbf{h}$ denotes the effective spin-splitting field in the model, representing the induced exchange magnetization. 
Experimentally, this quantity is indirectly tuned by the external magnetic field, $H$, via the hysteretic magnetization curve, and therefore is not expected to scale linearly with $H$. 
For a given set of parameters, the model yields the current--phase relation, from which we extract the positive and negative critical currents, $I_{\rm c}^\pm$, and use them to calculate $\eta$.

Following Ref.~\cite{Maiani2023}, we take homogeneous spin--orbit coupling, $\alpha=2~\mathrm{meV\,nm}$, and allow the spin-splitting field in the central region to rotate away from the wire axis by an angle $\theta_{\rm C}$, while the fields in the leads remain collinear with the wire. 
We fix the chemical potentials in the leads to $\mu_{\rm L}=\mu_{\rm R}=12\,\Delta_0$, where $\Delta$ is the superconducting gap, and vary only $\mu_{\rm C}$. 
We have checked that changing the values of $\mu_{\rm L}$ and $\mu_{\rm R}$, as well as spatially varying $\alpha_i$, gives qualitatively similar results.

The first basic result of the model is that a finite off-axis component of the spin-splitting field in the central region is required to obtain a finite JDE. 
This is illustrated in Fig.~\ref{f6}(b), which shows $\eta$ as a function of $\theta_{\rm C}$ at fixed $|h|=0.8\,\Delta_0$. 
As expected, $\eta$ vanishes at $\theta_{\rm C}=0$, where the spin-splitting fields in all three regions are collinear. 
Upon increasing $\theta_{\rm C}$, $\eta$ becomes finite and can change sign as the noncollinearity between the central region and the leads increases.

The broader evolution of the Josephson diode response is summarized by the phase diagram in Fig.~\ref{f6}(c), showing $\eta$ as a function of $\mu_{\rm C}$ and $h$ at fixed $\theta_{\rm C}=\pi/6$.
The calculation shows that $\eta$ vanishes near $h=0$, becomes finite as $h$ increases, and undergoes an abrupt sign change at a field value that depends on $\mu_{\rm C}$.
Within the model, this sign reversal occurs in the same parameter regime where the relative contribution of higher harmonics in the current--phase relation becomes more pronounced, consistent with the phase transitions discussed in Ref.~\cite{Maiani2023}.

We also find empirically that preserving the sign of $\eta$ upon reversal of the axial spin-splitting field requires the off-axis component in the central region to remain fixed. 
This is illustrated in Fig.~\ref{f6}(d), which compares $\eta(\mu_{\rm C})$ for different sign combinations of the axial and transverse components at fixed $|h|=0.8\,\Delta_0$ and $\theta_{\rm C}=\pi/6$. Within the model, reversing the axial component alone does not invert the sign of $\eta$, whereas reversing the off-axis component does. 
This suggests that the experimentally observed robustness of the diode polarity under axial magnetization reversal (see Appendix~\ref{app:D}) is consistent with a junction region in which the relevant off-axis component is governed by a fixed chirality, rather than by the sign of the axial magnetization alone. 
This provides a phenomenological condition for reproducing the sign-preserving behavior, but does not by itself establish a unique microscopic mechanism for the fixed off-axis component. 
We speculate that the magnetic texture is influenced by interface pinning~\cite{Diesch2018} and interactions~\cite{Kuepferling2023}.

Finally, to connect the model to the experimentally observed gate dependence, we consider selected $(\mu_{\rm C},\theta_{\rm C})$ pairs intended to mimic the effect of the back gate; see Fig.~\ref{f6}(e).
In the model, the back gate is represented primarily through the chemical potential in the central region, while possible gate-induced changes in the local spin--orbit environment are encoded phenomenologically by $\theta_{\rm C}$.
Increasing $\mu_{\rm C}$ and $\theta_{\rm C}$ leads to a larger $\eta$, in qualitative agreement with the experimental observation that $\eta$ increases with back-gate voltage.
In addition, within the selected parameter regime, the field scale at which the sign change occurs shifts toward lower $h$ as $\mu_{\rm C}$ and $\theta_{\rm C}$ increase. 
Combined with a sharp hysteretic magnetization curve---in which the effective spin splitting first deviates gradually from saturation and then switches rapidly at coercivity (see, for instance, Ref.~\cite{Razmadze2023})---this trend offers a natural explanation for the observed gate-dependent displacement of the sign change within the superconducting window.

\begin{figure}[t!]
 
    \centering
    \includegraphics[width = \linewidth]{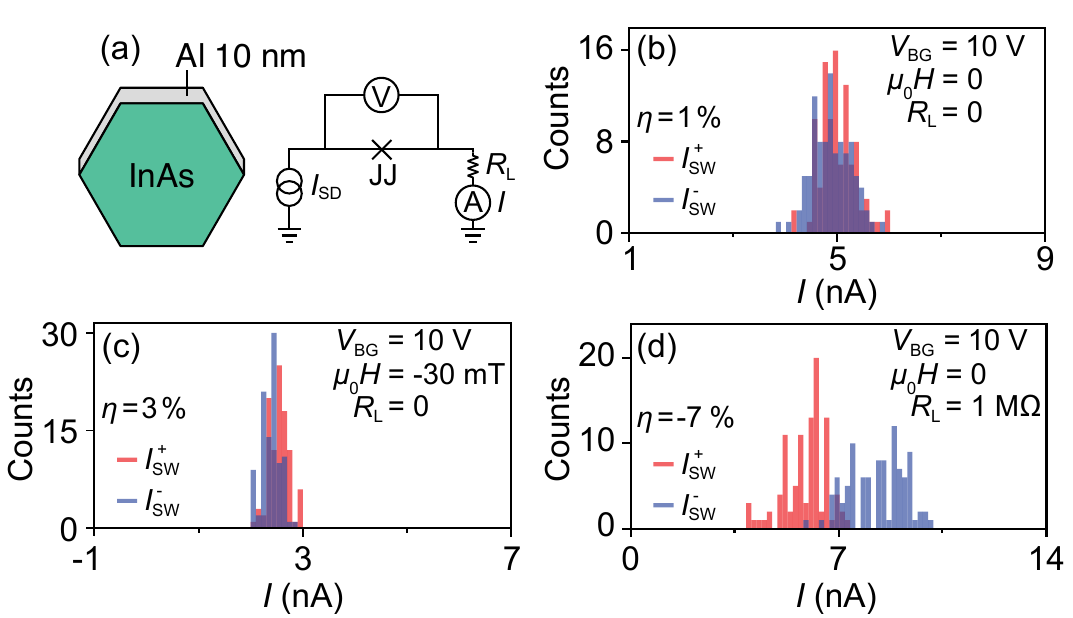}
    
    \caption{ 
    (a) Schematic cross-section of the non-magnetic Al/InAs nanowire used for control measurements together with the measurement setup. The load resistor, $R_{\rm L}$, was connected in series on the drain side only for the control measurements in (d).
    (b) Switching-current histograms at $V_{\mathrm{BG}}=10$~V and $\mu_0 H=0$, showing negligible JDE, $\eta=(1\pm0.5)\%$. 
    (c) Similar to (b) but at $\mu_0 H=-30$~mT, again with negligible $\eta=(3\pm0.5)\%$. 
    (d) Switching-current histograms measured with intentionally introduced asymmetric biasing through a load resistor $R_L=1~\mathrm{M}\Omega$, showing an apparent JDE, $\eta=(-7\pm0.3)\%$. 
    Data were taken at $V_{\rm JG}=-0.2$~V.
}
    \label{f7}
\end{figure}

\section{CONTROL MEASUREMENTS ON NON-MAGNETIC \NoCaseChange{Al/InAs} NANOWIRE}\label{app:H}

\subsection{Gate and Field Dependence}

To assess the role of the EuS shell in the observation of the JDE, we carried out control measurements on a reference Al/InAs nanowire without EuS; see Fig.~\ref{f7}(a). 
The junction was tuned to $V_{\mathrm{JG}} = -0.2$ V, such that it roughly matched the $I_{\rm SW}$ value observed for the ferromagnetic hybrid junctions. 
At zero applied magnetic field, $\mu_0 H = 0$, switching-current histograms measured at $V_{\mathrm{BG}}=+10$~V show statistically indistinguishable $I_{\rm SW}^+$ and $I_{\rm SW}^-$ distributions, with a negligible extracted Josephson diode efficiency of $\eta=(1\pm0.5)\%$; see Fig.~\ref{f7}(b). 
To test whether an external magnetic field applied parallel to the wire is by itself sufficient to induce nonreciprocity, we repeated the measurement at $\mu_0 H=-30$~mT under otherwise the same conditions. The extracted diode efficiency remains negligible, $\eta=(3\pm0.5)\%$; see Fig.~\ref{f7}(c). 
These control measurements show that neither gate bias alone nor a parallel magnetic field of this magnitude is sufficient to generate a measurable JDE in the absence of EuS. 

\subsection{Circuit artifacts}
We note that a diode-like response can, in principle, arise from an asymmetric resistive circuit that effectively shifts the device chemical potential by $e I R_{\rm Line}$ during a sweep~\cite{Shi2025}. In our setup, the only relevant series element on the drain side is a cryogenic filter with $R_{\rm Line}\lesssim 10$~k$\Omega$; together with $|I_{\rm SW}|\lesssim 10$~nA, this bounds the shift to $\lesssim 0.1$~meV. This is at least an order of magnitude below the meV-scale, gate-controlled chemical-potential variations typical for InAs hybrid nanowires~\cite{Antipov2018,Mikkelsen2018,Reeg2018,Escribano2021,Escribano2022}.

To experimentally test asymmetric circuit artifacts as a possible source of Josephson diode response, we performed a control experiment on the non-magnetic InAs/Al reference junction, in otherwise nominally the same device geometry, by intentionally introducing a load resistor $R_{\rm L}$ in series before the ground, similar to Ref.~\cite{Shi2025}. 
We find that reproducing a Josephson diode response comparable to that reported in Fig.~\ref{f3} requires $R_{\rm L}=1$~M$\Omega$ in an otherwise identical setup. 
This added resistance is more than two orders of magnitude larger than the fridge line resistance, $R_{\rm Line}$. 
In this configuration, we detect a sizable apparent diode efficiency, $\eta=(-7\pm0.3)\%$; see Fig.~\ref{f7}(d). 
We emphasize that without the added resistor, no JDE is observed in the same control device over a broad range of back-gate voltage at either zero or finite field; see Figs.~\ref{f7}(b) and \ref{f7}(c). The fact that a diode-like response appears only upon deliberately introducing such a large series resistance indicates that circuit-induced asymmetry is negligible in the main-text device. All measurements were performed in the same setup and manner as those discussed in the main text.

\begin{figure}[t!]
 
    \centering
    \includegraphics[width = \linewidth]{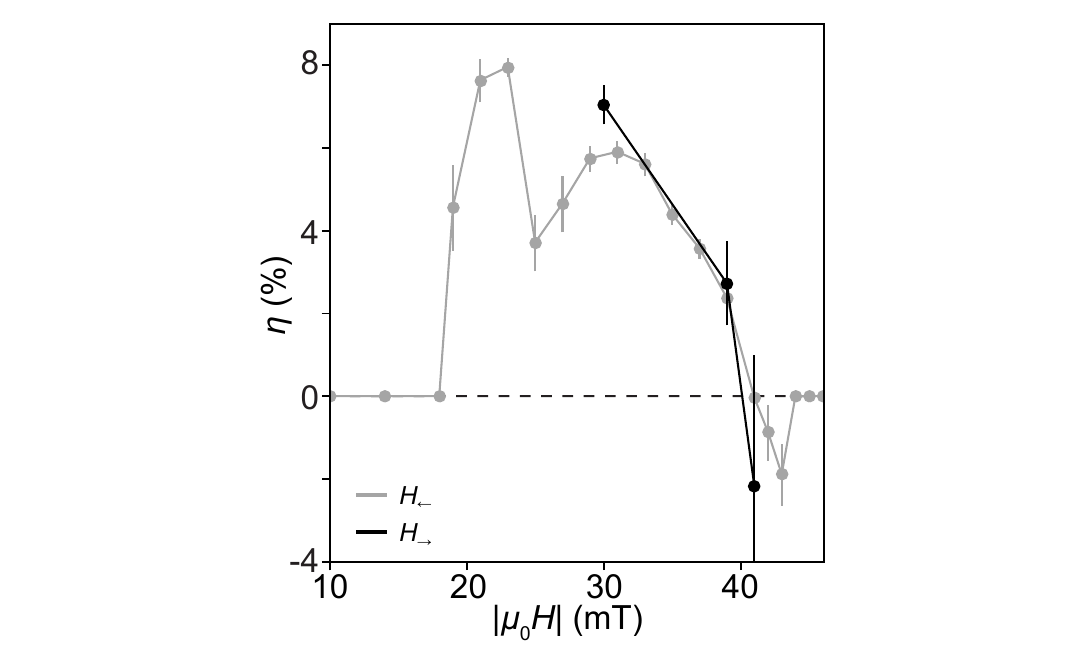}
    
    \caption{
    Josephson diode efficiency, $\eta$, as a function of magnetic field magnitude, $\vert\mu_0 H\vert$, obtained by sweeping $H$ from negative to positive (black) and from positive to negative (gray). The latter is the same data as in Fig.~\ref{f2}(c).
    These measurements confirm that the nonreciprocal response persists for both field sweep polarities, without an overall sign flip in $\eta$.
    Data were taken at $V_{\rm BG}=10$~V and $V_{\rm JG}=1.9$~V.}
    \label{f8}
\end{figure}

\begin{figure}[t!]
 
    \centering
    \includegraphics[width = \linewidth]{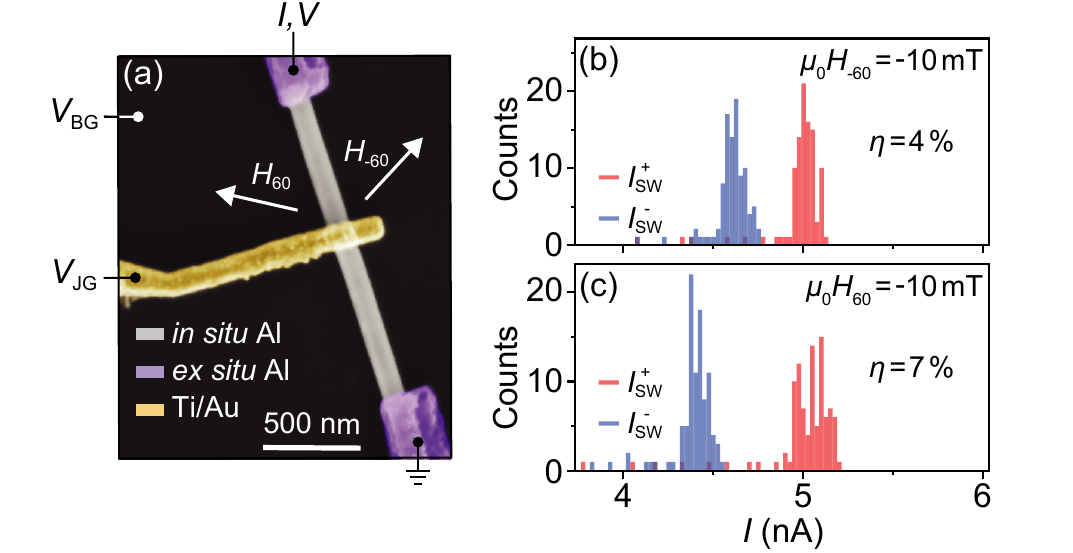}
    
    \caption{(a) Colorized scanning electron micrograph of the measured junction, with overlaid arrows indicating the applied magnetic field direction at angles of $\pm60^\circ$ relative to the nanowire axis.
    (b) and (c) Histograms of switching currents, $I_{\rm SW}^{+}$ and $I_{\rm SW}^{-}$, extracted from 100 repetitions at $\mu_0 H = -10$~mT with field applied at (b) $-60^\circ$ and (c) $+60^\circ$, after polarizing at $+100$~mT along the corresponding direction.
    The asymmetry in extracted Josephson diode efficiencies is consistent with the transverse field component adding to or subtracting from the intrinsic diode response.
    Data were taken at $V_{\rm BG}=10$~V and $V_{\rm JG}=1.9$~V.}
    \label{f9}
\end{figure}

\section{FIELD ORIENTATION DEPENDENCE}\label{app:D}

In the main text, we focus on the superconducting window on the negative magnetic field side.
To investigate the $H$-polarity dependence of the JDE, we performed analogous measurements at positive field.
We first polarized the EuS shell at $\mu_0 H = -100$~mT and then measured $I$--$V$ characteristics at positive field. 
The resulting Josephson diode efficiency, $\eta$, as a function of $H$, exhibits similar behavior to the negative-field case, without flipping the sign; see Fig.~\ref{f8}.

To investigate the angular dependence of JDE, we sweep the magnetic field at an angle of $\pm60^\circ$ relative to the wire axis; see Fig.~\ref{f9}. 
Switching current distributions at fixed field value $\mu_0 H_{\pm 60} = -10$~mT, where $\eta$ is near maximal, display clear separation between $I_{\mathrm{sw}}^+$ and $I_{\mathrm{sw}}^-$ for both $-60^\circ$ and $+60^\circ$ field orientations; see Figs.~\ref{f9}(b) and \ref{f9}(c).
The asymmetric efficiencies, $\eta = (4\pm0.2)\%$ for $-60^\circ$ and $\eta =(7\pm0.3)\%$ for $+60^\circ$, indicate that the perpendicular field component adds to or subtracts from the intrinsic Josephson diode response. 
The reduced $\eta$ compared to the parallel-field configuration may result from reduction of $I_{\rm SW}$ by the transverse field component and possible gate hysteresis between measurements.

\begin{figure}[t!]
    \centering
    \includegraphics[width = \linewidth]{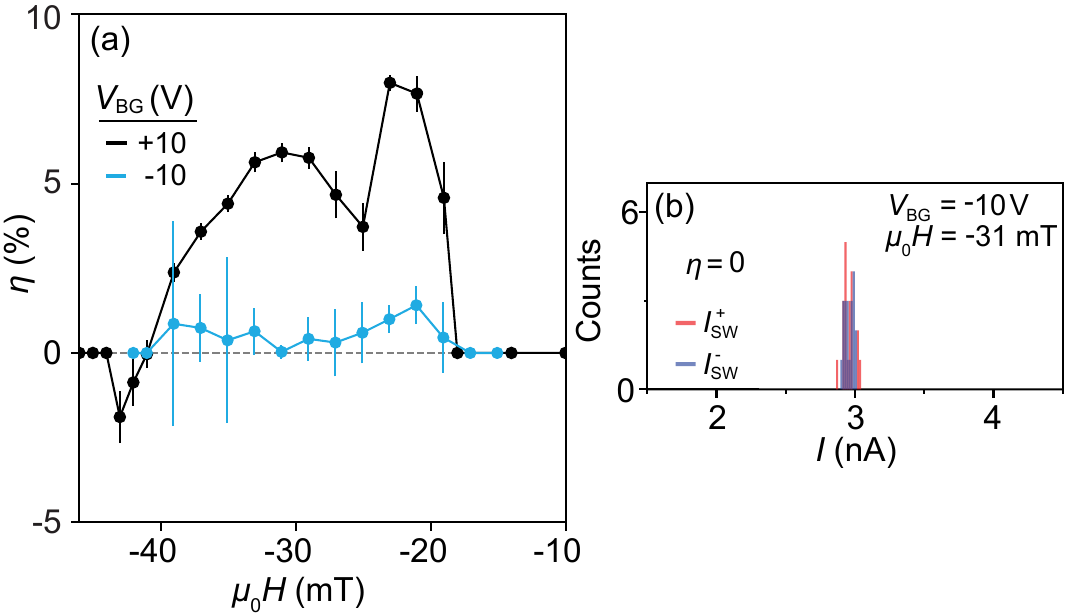}
    \caption{
    (a)~Josephson diode efficiency, $\eta$, as a function of magnetic field, $H$, for $V_{\rm BG}=+10$~V [same data as in Fig.~\ref{f2}(c)] and $V_{\rm BG}=-10$~V. 
    While a finite diode response is observed at positive $V_{\rm BG}$, the extracted efficiency at $V_{\mathrm{BG}}=-10$~V remains predominantly consistent with zero within experimental uncertainty throughout the superconducting window.
    (b) Representative switching-current distributions at $V_{\rm BG}=-10$~V and $\mu_0H=-31$~mT, yielding $\eta=(0.0 \pm 0.2)\%$, consistent with zero. 
    Data were taken at $V_{\rm JG}=1.9$~V.
    }
    \label{f10}
\end{figure}

\section{GATE-VOLTAGE DEPENDENCE}\label{app:F}

\subsection{Extended back-gate dependence}

As an additional check, we further examine the asymmetric back-gate dependence of the diode response.
At $V_{\rm BG}=-10$~V, $\eta$ remains predominantly consistent with zero throughout the superconducting window, in contrast to the finite response observed at $V_{\rm BG}=+10$~V; see Fig.~\ref{f10}(a).
A representative switching-current histogram at $V_{\rm BG}=-10$~V and $\mu_0H=-31$~mT is shown in Fig.~\ref{f10}(b), where the positive- and negative-current distributions overlap and yield $\eta=(0.0 \pm 0.2)\%$, consistent with zero.
Together with the gate dependence presented in Fig.~\ref{f3}, these data show that the diode response is strongly asymmetric with respect to $V_{\rm BG}=0$: it is suppressed for negative back-gate voltage and grows systematically for positive $V_{\rm BG}$.
This behavior is inconsistent with a leakage-induced current offset and is instead a reproducible feature of the junction response.

\begin{figure}[!t]
 
    \centering
    \includegraphics[width = \linewidth]{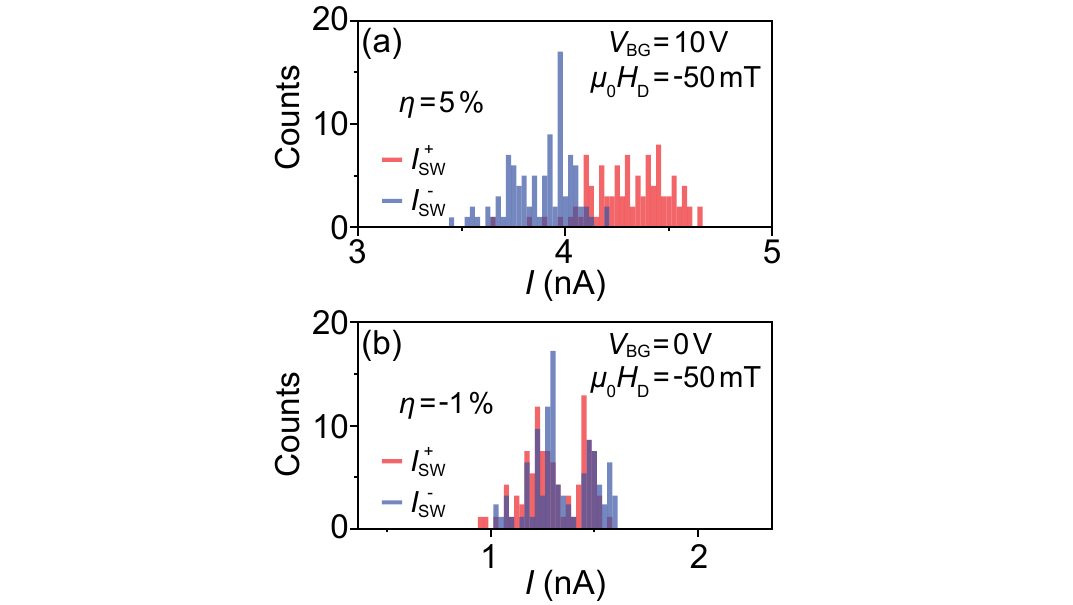}
    
    \caption{(a) Histograms of switching currents, $I_{\mathrm{sw}}^{+}$ and $I_{\mathrm{sw}}^{-}$, measured at $H=0$, after applying demagnetizing field of $\mu_0H_{\rm{D}} = -50$~mT and back-gate voltage $V_{\rm BG} = 10$~V.
    The two histograms are distinct, giving a sizable Josephson diode efficiency of $\eta=(5\pm 0.4)\%$.
    (b) Same as (a) but for $V_{\rm BG} = 0$, showing nearly indistinguishable histograms with $\eta=(-1\pm 0.8)\%$.
    Each histogram is compiled from 100 current sweeps in both directions.
    Data were taken at $V_{\rm JG}=1.9$~V.}
    \label{f11}
\end{figure}

\begin{figure}[t!]
 
    \centering
    \includegraphics[width = \linewidth]{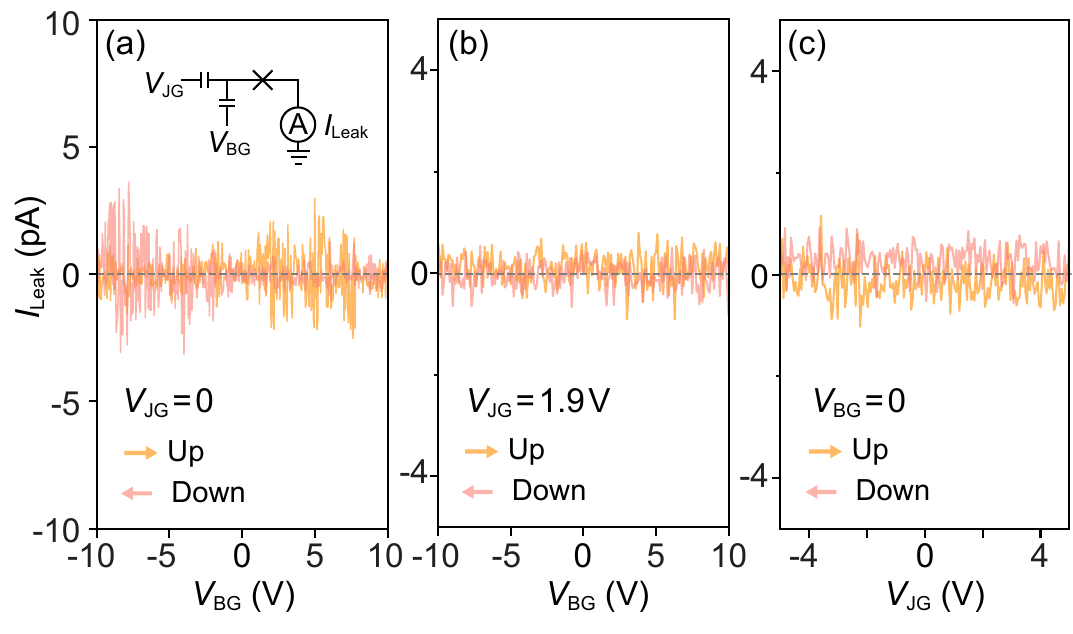}
    
    \caption{
    (a) Leakage current, $I_{\mathrm{Leak}}$, measured for the main junction as a function of back-gate voltage, $V_{\rm BG}$, for upward and downward sweeps at $V_{\rm JG}=0$. 
    (b) Same as (a) but for $V_{\rm JG}=1.9$~V, corresponding to the junction-gate setting used in the main experiment.
    (c) Leakage current as a function of junction-gate voltage, $V_{\rm JG}$, at fixed $V_{\rm BG}=0$.
    No measurable leakage current is detected above the experimental noise floor over the relevant gate-voltage ranges.
    These data were taken in a separate cooldown at $T=7$~K in a different cryostat. 
    }
    \label{f12}
\end{figure}

\begin{figure}[t!]
 
    \centering
    \includegraphics[width =\linewidth]{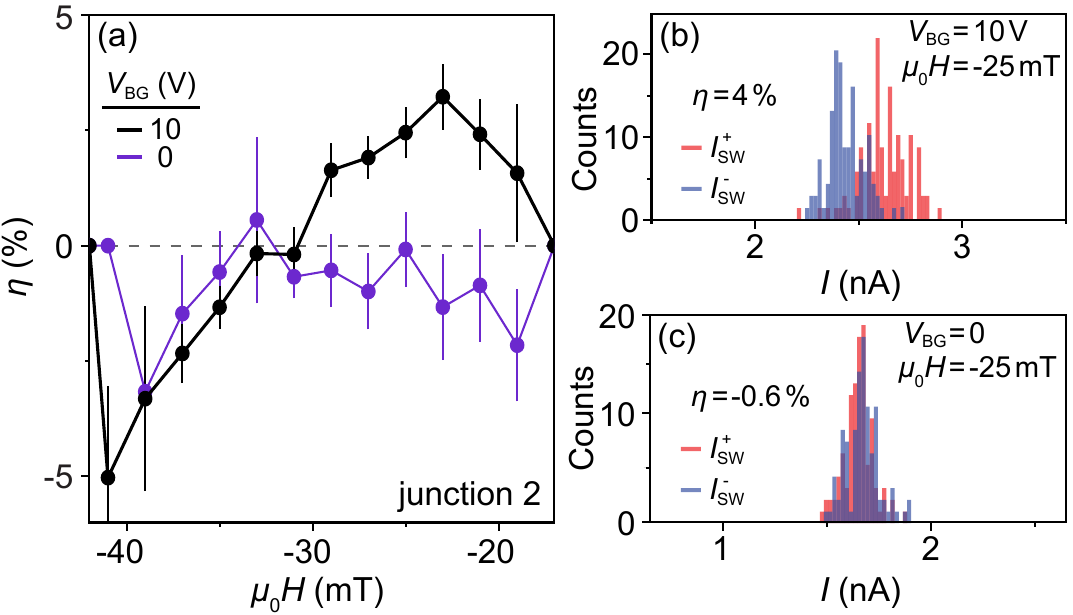}
    
    \caption{ 
    (a) Josephson diode efficiency, $\eta$, as a function of magnetic field, $H$, measured for junction 2 (nominally identical to junction 1 reported in the main text) at back-gate voltages $V_{\rm BG} = 10$ and 0~V. Each point is calculated from 20 current-sweep cycles.
    (b) and (c) Histograms of switching currents, $I_{\rm SW}^{+}$ and $I_{\rm SW}^{-}$, extracted from separate 100 repetitions at $\mu_0 H = -25$~mT for (b) $V_{\rm BG} = 10$~V, with sizable $\eta = (4 \pm 0.3)\%$, and (c) $V_{\rm BG} = 0$, where $\eta = (-0.6 \pm 0.3)\%$ is consistent with zero.
    Data were taken at $V_{\rm JG}=2$~V.}
    \label{f13}
\end{figure}

\subsection{Voltage control at $H=0$}

To investigate whether JDE remains gate-tunable in the absence of an external magnetic field, we measured switching current histograms at $H = 0$ following a demagnetizing field of $\mu_0 H_{\rm D} = -50$~mT, for two different back-gate voltages; see Fig.~\ref{f11}.
We find a sizable Josephson diode efficiency of $\eta = (5 \pm 0.4)\%$ at $V_{\rm BG} = 10$~V, and a suppressed efficiency of $\eta = (-1 \pm 0.8)\%$, consistent with zero, at $V_{\rm BG} = 0$.
These measurements confirm that gate control over nonreciprocal supercurrent transport persists in the demagnetized regime, consistent with the behavior observed under finite applied fields; see Fig.~\ref{f3}.

We note that the residual field after the magnet reset is estimated to be at most a few millitesla. Such a residual field or trapped flux would act primarily as an additional field offset. These effects cannot, however, account for the pronounced back-gate dependence observed under the same magnetic-field preparation; see Fig.~\ref{f11}.

\subsection{Gate-leakage tests}

To rule out gate leakage as a possible origin of the observed nonreciprocity, we measured the leakage current, $I_{\rm Leak}$, under the relevant gate configurations. 
Figure~\ref{f12}(a) shows $I_{\rm Leak}$ as a function of back-gate voltage, $V_{\rm BG}$, for $V_{\rm JG}=0$, while Fig.~\ref{f12}(b) shows the corresponding measurement for the junction-gate setting used in the main experiment, $V_{\rm JG}=1.9$~V. 
We additionally measure the leakage current while sweeping the junction gate, $V_{\rm JG}$, at fixed $V_{\rm BG}=0$; see Fig.~\ref{f12}(c).
In all cases, no measurable leakage current is detected above the experimental noise floor over the relevant operating range. 
The noise level is roughly two orders of magnitude smaller than the typical measured difference between $I_{\mathrm{SW}}^{+}$ and $I_{\mathrm{SW}}^{-}$. 
Moreover, at fixed gate voltage, a leakage-induced current offset would be expected to remain approximately constant and magnetic-field independent, which is inconsistent with the pronounced field dependence and sign reversal of the Josephson diode efficiency, $\eta$, observed experimentally; see Fig.~\ref{f2}. 

\section{SECOND JUNCTION CHARACTERIZATION}\label{app:G}

We measured a second Josephson junction, nominally identical to that reported in the main text, to demonstrate reproducibility. 
Josephson diode efficiency, $\eta$, as a function of magnetic field, $H$, measured for $V_{\mathrm{BG}} = 0$ and $10$~V, shows a gate-voltage dependence consistent with the main device; see Fig.~\ref{f13}(a).
Switching current histograms at $\mu_0 H = -25$~mT confirm enhanced nonreciprocity at more positive $V_{\rm BG}$, with $\eta = (4 \pm 0.3)\%$ for $V_{\rm BG} = 10$~V and negligible $\eta = (-0.6 \pm 0.3)\%$ for $V_{\rm BG} = 0$; see Figs.~\ref{f13}(b) and \ref{f13}(c).
Compared to the main device, this junction exhibits slightly lower $I_{\rm SW}$ and $\eta$, which we attribute to fabrication-induced variation.

\bibliography{papers.bib}

\end{document}